# Too Sick for Working, or Sick of Working?
## Impact of Acute Health Shocks on Early Labour Market Exits

Luis Vieira

*Massey Business School*
*Massey University*
*Auckland, New Zealand*
23012096@massey.ac.nz

**Abstract**. This study investigates the impact of acute health shocks on early labour market exits among individuals aged 50 and over in Europe. Utilising data from Waves 1-8 of the Survey of Health, Ageing, and Retirement in Europe (SHARE), encompassing over 140,000 individuals, I employ survival analysis techniques, including Kaplan-Meier estimators and Cox proportional hazards models. The analysis explores how sudden, severe health events, alongside self-perceived health, chronic conditions, and socio-demographic factors, influence the probability of exiting the workforce before official retirement age. Results indicate that acute health shocks significantly increase the hazard of early labour market exit, with more pronounced effects observed for males. Poorer self-perceived health and lower educational attainment are also strong predictors of early exit. Gender differences are notable: while poorer health consistently raises exit risk for both genders, income acts as a protective factor for females, and living with a partner reduces exit risk for males but increases it for females. These findings highlight the critical role of health in labour market participation and suggest the need for targeted policies to support older workers, particularly those experiencing adverse health events.

## 1 Introduction

The concept of retirement, as we know it today, dates back to 1881 when Otto von Bismarck, former Chancellor of Germany, introduced a national retirement scheme with the retirement age set at 70. At the time, the average life expectancy was approximately 40 years, so few ever reached retirement age. This threshold was later adjusted to 65, a figure that has persisted for over a century. However, today's rapidly ageing population, coupled with the rising costs of welfare, has brought renewed reforms to retirement policies and the economic sustainability of ageing societies.

Understanding the drivers behind early retirement is critical, particularly the role of health. This study investigates the impact of acute health shocks on early labour market exits, leveraging the Survey of Health, Ageing, and Retirement in Europe (SHARE), which provides a wealth of data from over 140,000 individuals aged 50 and above across 28 European countries and Israel. Health shocks, which are defined as sudden and severe declines in health status, have been shown to have profound effects on individuals' work trajectories, pushing many towards early retirement and imposing financial strain on both the individuals and the welfare system (Schofield et al., 2011).

Research has consistently shown that health significantly influences labour market participation. García-Gómez, Jones, and Rice (2008) found that health limitations and mental health issues increase the hazard of becoming non-employed, with stronger effects for men, using data from the British Household Panel Survey. Macchiarelli and Aranki (2013) highlighted that in 26 EU countries, income and flexible working conditions are critical in early retirement decisions, especially during economic crises, but state and health benefits alone are insufficient if health deteriorates. Zucchelli et al. (2010), using the HILDA survey, reported that acute health shocks raise the risk of labour market exit, with the hazard increasing between 50% and 320% for men and between 68% and 74% for women. In short, these studies highlight the complex impacts of health, financial resources, and institutional policies in influencing labour market outcomes.

From a methodological point, survival analysis offers a robust framework for examining these transitions. The Kaplan-Meier estimator (Kaplan & Meier, 1958) provides a nonparametric approach to estimating the duration until an event, such as labour market exit, while the Cox proportional hazards model (Cox, 1972) allows for an assessment of how various covariates – such as acute and other health shocks, chronic conditions, and demographic factors – influence the risk of leaving the workforce. These models have been widely used in the literature, with Harrell et al. (1982) and Antolini et al. (2005) emphasising the importance of model performance metrics, such as the concordance index and its time-dependent variants, to ensure reliable predictions.

Acute health shocks, such as heart attacks, strokes, and severe fractures, often lead to long-term medical care and lifestyle adjustments, which can be decisive in an individual's decision to exit the labour market early. The financial repercussions of these health shocks can be severe, as individuals who retire early due to health reasons typically have lower lifetime earnings, less accumulated wealth, and may face higher healthcare costs (Schofield et al., 2011). Moreover, retirement decisions are not made in isolation. Adequate pension benefits may incentivise retirement, whereas insufficient savings may force individuals to remain employed. Psychologically, job dissatisfaction or the idea of leisure may push workers towards retirement, while social factors like caregiving responsibilities or family dynamics can also be influential.

This research aims to shed light on these complex dynamics, focusing on how acute health shocks and self-perceived health status influence early retirement decisions across European countries. By employing survival analysis techniques, this study aims to contribute by providing a nuanced understanding of the factors driving early exits from the workforce, with a focus on acute health events and their broader implications for welfare policy and financial security.

## 2 Data & Descriptive Statistics

### 2.1 Data Description

The analysis utilises a subset of the European Survey of Health, Ageing and Retirement in Europe (SHARE) dataset. "SHARE is a cross-national panel dataset of more than 140,000 subjects, aged 50 and over, from 28 European countries and Israel. The dataset provides observations on health, socio-economic status and social and family networks" (Börsch-Supan, 2013). Following individuals over waves 1 to 8, the panel data's individuals aged between 50 and their retirement age, data was systematically filtered and cleaned to address missing values and ensure robustness. The dataset captures a diverse range of socioeconomic, demographic, and health-related variables essential for modelling early labour market exits, especially in response to acute health shocks.

The data structure facilitates time-to-event analysis, where the primary outcome is the early exit from the labour market, measured using a survival analysis framework. Below, the detail of main variables used in the analysis (Table 1).

Table 1 – Variables Used for Analysing Early Labour Exits

| Variables | Description |
|---|---|
| *Dependent variable* | |
| Labour force exit | 1 if the respondent exited the labour force before retirement, 0 otherwise |
| *Health measures* | |
| Self-Perceived Health (SPH) | Self-assessed health status: 1 = excellent, 2 = very good, 3 = good, 4 = fair, 5 = poor |
| SPH binary | 1 if self-perceived health is fair or poor (4–5), 0 if good, very good, or excellent (1–3) |
| Chronic diseases (chronic_d) | 1 if the respondent has one or more chronic illnesses, 0 otherwise |
| Low mobility | 1 if the respondent has severe physical limitations, 0 otherwise |
| *Health shocks* | |
| Acute health shock | 1 if suffered a severe health event (e.g., heart attack, stroke) in the past 12 months, 0 otherwise |
| Negative health shocks | 1 if experienced any significant health deterioration across survey waves, 0 otherwise |
| Psychological shocks | 1 if experienced cumulative psychological distress events, 0 otherwise |
| *Household variables* | |
| Household net income (thinc_m) | Total household income before transformation (in thousand Euros) |
| Log household income | Logarithm of total household income, adjusted for skewness |
| Financial strain (ends_notmet) | 1 if ends meet with difficulty or barely, 0 otherwise |
| Household composition (hh_comp) | Categorised as solo, two-person, or multi-person household |
| Living with partner (partner_hh) | 1 if living with a partner, 0 otherwise |
| *Socioeconomic variables* | |
| Employment status | Categorical: 1 = retired, 2 = employed/self-employed, 3 = unemployed, 4 = permanently sick/disabled, 5 = homemaker |
| Education level | Categorical: 1 = university degree, 2 = diploma/certificate, 3 = secondary education, 4 = basic education, 5 = other |
| *Demographics* | |
| Age | Age of respondent in each wave |
| Gender (female) | 1 if female, 0 if male |
| Country of residence (country_name) | Nominal variable specifying the respondent's country of residence for comparative analysis |
| Retirement age | Official retirement age based on country and gender |

## 2.2 Data Preparation & Transformations

The data used in this study required extensive cleaning and transformation to ensure it was adequately structured for survival analysis. The initial step involved handling missing values, where negative and implausible entries across key variables – such as health indicators (ph006d1-ph006d21), financial measures (ends_meet, thinc_m), or mobility limitations (mobilityind) – were recoded to missing.

Country codes, were translated into descriptive labels using the variable country_name. This transformation facilitated clearer interpretation and readability in cross-country comparisons. Health measures were meticulously constructed, with cumulative negative health shocks (neg_HShocks) reflecting aggregated adverse health events experienced across survey waves. Acute health shocks (acute_HShock), defined as sudden health events like heart attacks, strokes or severe fractures reported since the last interview, provided critical insights into the immediate effects of severe health deterioration on labour market exits. Additionally, cumulative psychological shocks (neg_PsyShocks) captured the accumulation of significant psychological stressors, which could indirectly affect employment status.

Economic variables were transformed to address issues of skewness. Household net income (thinc_m), originally reported in thousands of Euros, was adjusted to Euros and then log-transformed to produce log_inc_m, avoiding negative log values when incomes were less than 1 (1000 Euros). This transformation enabled a more symmetrical income distribution, enhancing interpretive clarity. Financial strain, derived from the ends_meet variable and captured by ends_notmet, was formulated as a binary indicator highlighting economic vulnerability among respondents struggling to make ends meet.

Categorical variables were constructed to encapsulate a range of socioeconomic and demographic attributes. Employment status (employ_status) was reclassified into categories such as employed, unemployed, retired, permanently sick or disabled, and homemaker. The dependent variable exit_labourforce was defined to indicate labour market activity, distinguishing active participants from those who had exited due to various circumstances, such as retirement or health issues. Chronic conditions (chronic_d) and low mobility (low_mobility) were also represented as binary indicators, derived from the presence of chronic illnesses and substantial physical limitations, respectively.

Education level was operationalised as an ordinal variable, education_level, categorising respondents based on their highest completed education: University Degree, Diploma/Certificate, Secondary Education, Basic Education, and Other Education. This classification allowed for nuanced analysis of how educational attainment influences labour market behaviour. Household composition (hh_comp) categorised households into solo, two-person, or multi-person households, reflecting potential familial dependencies and financial impact of cohabitation. The variable partner_hh captured whether an individual lived with a partner, providing insight into the possible social and financial support.

Self-perceived health, a critical component in the analysis, was represented by the variable SPH, renamed from sphus and organised on a Likert scale from Excellent to Poor. An additional binary variable, SPH_binary, was created to simplify health status into Good (Excellent-Good) and Poor (Fair-Poor) categories, facilitating clearer statistical comparisons.

Retirement age, was constructed using retirement_age, which varies by country and gender (Figure 1). This variable was crucial for defining the censoring thresholds in the survival models and understanding early exits from the labour force. Note that retirement age was rounded up in the cases where it was not a round year.

Figure 1 - Official Retirement Age by Country and Gender (2023)

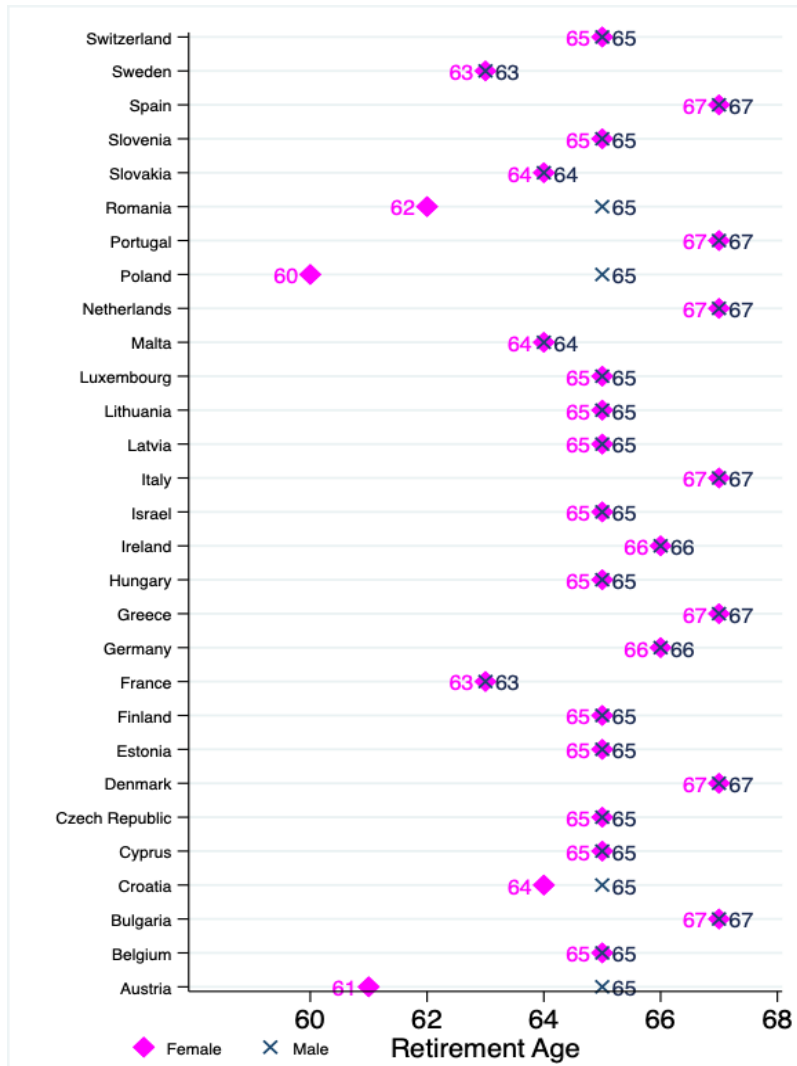

*Note*. Adapted from Finnish Centre for Pensions

**Survival Analysis Methodology**

The survival analysis models time-to-event data, where the event of interest is early retirement or labour force exit. This approach captures the timing of labour market exits among individuals aged 50 and older, providing insights into how health and other factors influence these transitions.

Managing right-censoring is important, as it accounts for individuals who remain employed at the study's conclusion. By incorporating right-censoring, the analysis retains the full sample and avoids bias, with the discrete-time hazard model handling these cases. The data is also subject to left truncation, the analysis focuses on individuals aged 50 or over and up to one year before retirement age at their initial wave, ensuring that only those genuinely at risk of early labour force exit are included.

Event and time variables were carefully defined: the event variable indicates labour market exit, with t0 and t1 representing the ages at entry and exit. Adjustments for right-censoring were made using censor_age to ensure accurate representation of individuals who reached retirement age. Additionally, cases of panel attrition – where individuals left the study before labour market exit – were identified and excluded to maintain data integrity, ensuring that attrition did not skew the results. The variable country_name was encoded into country_id to facilitate analysis, and all variables were systematically labelled. Using the stset command in Stata, the dataset was configured for survival analysis, with t1_adj as the time-to-event variable, t0 as the entry age, and event as the indicator of labour force exit. This rigorous setup accounts for right-censoring, left truncation, and attrition, ensuring a robust examination of health shocks and other influences on early exits from the labour market.

## 2.3 Descriptive Statistics

Table 2 provides a detailed overview of the health, socioeconomic, and demographic characteristics of the sample, broken down by gender and labour market status. This comprehensive summary provides a basis on the factors influencing early labour market exits, forming the basis for our survival analysis.

Table 2 - Descriptive Statistics

|  | Men | | | Women | | |
|---|---|---|---|---|---|---|
|  | All | Active | Inactive | All | Active | Inactive |
| *Health & Psychologic Variables* | | | | | | |
| Self-perceived health (SPH) | | | | | | |
| SPH Excellent | 0.095 | 0.117 | 0.060 | 0.090 | 0.118 | 0.061 |
| SPH Very Good | 0.227 | 0.266 | 0.166 | 0.218 | 0.260 | 0.175 |
| SPH Good | 0.387 | 0.402 | 0.363 | 0.392 | 0.400 | 0.383 |
| SPH Fair | 0.220 | 0.183 | 0.279 | 0.237 | 0.193 | 0.283 |
| SPH Poor | 0.071 | 0.031 | 0.133 | 0.064 | 0.029 | 0.099 |
| Cumulative negative health shocks | 0.676 | 0.606 | 0.786 | 0.689 | 0.614 | 0.767 |
| Acute health shock | | | | | | |
| No New Shock =0 | 0.974 | 0.984 | 0.958 | 0.985 | 0.991 | 0.979 |
| New Shock =1 | 0.026 | 0.016 | 0.042 | 0.015 | 0.009 | 0.022 |
| Cumulative psychological shocks | 0.542 | 0.520 | 0.577 | 0.630 | 0.611 | 0.650 |
| Has chronic diseases | | | | | | |
| No =0 | 0.452 | 0.521 | 0.343 | 0.494 | 0.576 | 0.411 |
| Yes =1 | 0.548 | 0.479 | 0.657 | 0.506 | 0.424 | 0.589 |
| Has low mobility | | | | | | |
| No =0 | 0.970 | 0.990 | 0.938 | 0.968 | 0.987 | 0.949 |
| Yes =1 | 0.030 | 0.010 | 0.062 | 0.032 | 0.013 | 0.051 |
| *Demographic & Socioeconomic Variables* | | | | | | |
| Age | 60.1 | 58.8 | 62.1 | 59.7 | 58.1 | 61.3 |
| Living with Partner | | | | | | |
| No =0 | 0.170 | 0.162 | 0.182 | 0.235 | 0.252 | 0.217 |
| Yes =1 | 0.830 | 0.838 | 0.818 | 0.765 | 0.748 | 0.783 |
| Household composition | | | | | | |
| Solo Household | 0.127 | 0.120 | 0.138 | 0.159 | 0.169 | 0.149 |
| Two-person Household | 0.520 | 0.485 | 0.574 | 0.557 | 0.537 | 0.577 |
| Three or more person Household | 0.353 | 0.394 | 0.287 | 0.285 | 0.294 | 0.275 |
| Monthly household income (Euros) | 36,139 | 38,931 | 31,940 | 32,800 | 37,445 | 28,171 |
| Log of monthly household income | 16.978 | 17.043 | 16.879 | 16.896 | 17.053 | 16.739 |
| Level of education | | | | | | |
| University Degree | 0.256 | 0.301 | 0.185 | 0.254 | 0.343 | 0.162 |
| Diploma/Certificate | 0.047 | 0.055 | 0.034 | 0.047 | 0.061 | 0.032 |
| Secondary Education | 0.398 | 0.398 | 0.398 | 0.362 | 0.374 | 0.350 |
| Basic Education | 0.288 | 0.236 | 0.370 | 0.327 | 0.212 | 0.445 |
| Other Education | 0.011 | 0.009 | 0.013 | 0.011 | 0.010 | 0.016 |
| Employment Status | | | | | | |
| Retired | 0.311 | . | 0.810 | 0.262 | . | 0.531 |
| Employed or self-employed | 0.552 | 0.902 | . | 0.461 | 0.911 | . |
| Unemployed | 0.060 | 0.098 | . | 0.045 | 0.089 | . |
| Permanently sick or disabled | 0.069 | . | 0.177 | 0.059 | . | 0.119 |
| Homemaker | 0.005 | . | 0.013 | 0.173 | . | 0.350 |
| Country | | | | | | |
| Austria | 0.051 | 0.042 | 0.064 | 0.032 | 0.033 | 0.031 |
| Belgium | 0.104 | 0.095 | 0.118 | 0.095 | 0.095 | 0.095 |
| Bulgaria | 0.004 | 0.004 | 0.003 | 0.004 | 0.004 | 0.004 |
| Croatia | 0.015 | 0.013 | 0.017 | 0.014 | 0.013 | 0.016 |
| Cyprus | 0.001 | 0.001 | 0.001 | 0.001 | 0.001 | 0.001 |
| Czech Republic | 0.057 | 0.051 | 0.066 | 0.068 | 0.050 | 0.086 |
| Denmark | 0.083 | 0.099 | 0.056 | 0.076 | 0.095 | 0.056 |
| Estonia | 0.066 | 0.072 | 0.058 | 0.075 | 0.101 | 0.049 |
| Finland | 0.005 | 0.005 | 0.004 | 0.004 | 0.006 | 0.002 |
| France | 0.062 | 0.062 | 0.062 | 0.061 | 0.072 | 0.049 |
| Germany | 0.074 | 0.080 | 0.063 | 0.076 | 0.089 | 0.063 |
| Greece | 0.046 | 0.048 | 0.042 | 0.049 | 0.030 | 0.070 |
| Hungary | 0.007 | 0.006 | 0.009 | 0.008 | 0.005 | 0.011 |
| Israel | 0.026 | 0.032 | 0.018 | 0.033 | 0.032 | 0.033 |
| Italy | 0.080 | 0.071 | 0.094 | 0.086 | 0.054 | 0.118 |
| Latvia | 0.003 | 0.004 | 0.003 | 0.003 | 0.004 | 0.003 |
| Lithuania | 0.005 | 0.007 | 0.004 | 0.007 | 0.010 | 0.004 |
| Luxembourg | 0.016 | 0.010 | 0.024 | 0.017 | 0.012 | 0.023 |
| Malta | 0.003 | 0.003 | 0.002 | 0.003 | 0.002 | 0.004 |
| Netherlands | 0.040 | 0.036 | 0.046 | 0.043 | 0.036 | 0.051 |
| Poland | 0.026 | 0.023 | 0.030 | 0.012 | 0.015 | 0.009 |
| Portugal | 0.011 | 0.008 | 0.016 | 0.013 | 0.010 | 0.015 |
| Romania | 0.005 | 0.003 | 0.009 | 0.004 | 0.003 | 0.006 |
| Slovakia | 0.006 | 0.007 | 0.004 | 0.005 | 0.008 | 0.003 |
| Slovenia | 0.042 | 0.030 | 0.061 | 0.047 | 0.032 | 0.062 |
| Spain | 0.082 | 0.076 | 0.092 | 0.080 | 0.065 | 0.096 |
| Sweden | 0.034 | 0.048 | 0.011 | 0.034 | 0.056 | 0.012 |
| Switzerland | 0.049 | 0.065 | 0.024 | 0.049 | 0.067 | 0.031 |

**Note**: Ireland had no observations with Acute Health Shocks and labour market exits.

Total Obs: 85,156

The health and psychological variables reveal considerable differences between active and inactive individuals, but only small difference between gender. Self-perceived health (SPH) shows that inactive participants report substantially poorer health. Among men, 41.2% of those who are inactive rate their health as "Fair" or "Poor," compared to only 21.4% of the active group. For women, 38.2% of the inactive cohort reports "Fair" or "Poor" health, compared to 22.2% of active women. Acute health shocks are more prevalent among those who have exited the labour force: 4.2% of inactive men and 2.2% of inactive women experienced a severe health shock since the last interview, compared to just 1.6% and 0.9% of active men and women, respectively. Chronic illnesses and low mobility are also higher among inactive individuals, with 65.7% of inactive men and 58.9% of inactive women suffering from chronic conditions.

Demographic and socioeconomic differences are also evident. Age shows inactive men averaging 62.1 years and inactive women 61.3 years, both clearly under the retirement age. Household composition data show that inactive women are more likely to live alone (14.9%) than active women (12.0%), suggesting the importance of social and economic support networks. Economic disparities are clear, as inactive men and women report significantly lower average monthly household incomes (€31,940 for men and €28,171 for women) compared to the active (€38,931 for men and €37,445 for women). Education levels show that inactive individuals more often have only basic education, potentially linking lower educational attainment with earlier labour force withdrawal.

Most countries had relatively small sample sizes, making it challenging to draw country-specific conclusions. However, Belgium, Italy, and Spain stood out with over 1,000 exits/events for both genders. These countries may warrant closer examination in further analysis, given their higher representation and potential influence on overall trends.

Figure 2 - Proportion of Labour Exits by Age and Gender (in percentage)

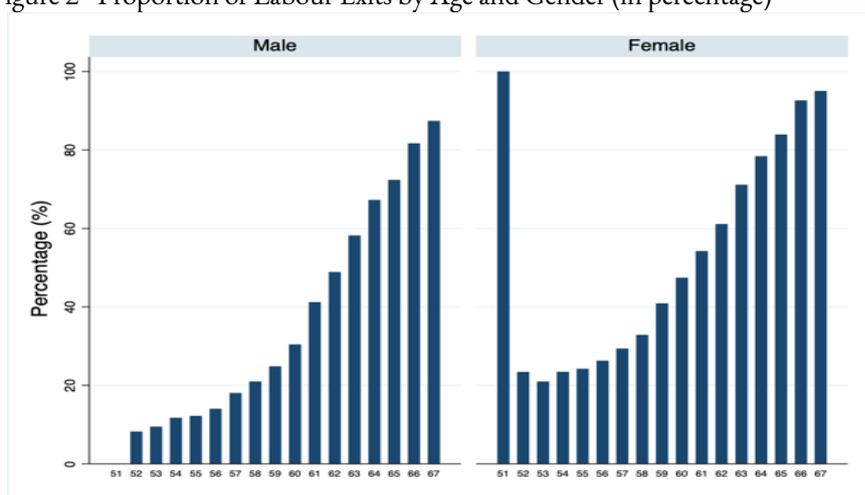

The bar chart in Figure 2 reveal age-related patterns in labour market exits, with a steep increase in exit rates around 60. Women generally exit the labour force at a younger age compared to men (note also the outlier female exit percentage at age 51), likely influenced by not only the commonly earlier retirement ages but also caregiving responsibilities or health-related factors.

Figure 3 – Acute Health Shocks Occurrences by Age (in percentage)

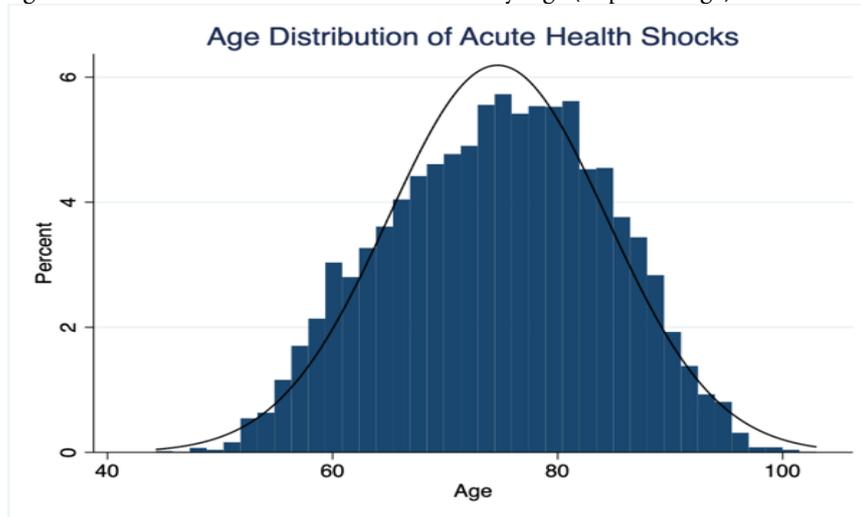

The histogram in Figure 3 shows that acute health shocks also become more frequent with age, peaking around the early 70s. This trend highlights the need to consider age-specific health risks when modelling labour market exits and planning for age-related employment policies.

These observations illustrate the complex relationship between health status, age, and socioeconomic conditions in influencing early labour market exits. Health challenges, economic vulnerabilities, and lower educational are likely central factors driving early withdrawal.

## 3 Survival Analysis

The survival analysis uses the Kaplan-Meier estimator to assess the probability of remaining in the labour force over time, stratified by SPH, gender, and the occurrence of acute health shocks. This non-parametric approach provides a robust overview of survival differences between groups, allowing a clear visual representation of how health and demographic factors affect labour force retention as individuals approach retirement age. Additionally, a Cox proportional hazards model was later used to quantify the impact of various covariates on the hazard of exiting the labour market.

### 3.1 Kaplan-Meier Survival Estimates

The Kaplan-Meier survival analysis in Figure 4 highlights gender differences in labour market exits in response to acute health shocks. By age 60, about 75% of males (navy line) without acute health shocks remain in the labour market, compared to only around 55% of males (dashed navy) who have experienced such shocks, indicating a pronounced impact. For females, around 55% without health shocks (in purple) are still active at age 60, declining to 40% for those with health shocks (dashed purple). Although women generally retire earlier and also due to lower statutory retirement ages, the effect of acute health shocks is more pronounced among males. This suggests that men experience a steeper decline in labour market participation following severe health events.

Figure 4 – Kaplan-Meier Survival Estimates of the Proportion of Labour Market Exits by Acute Health Shocks and Gender (all countries)

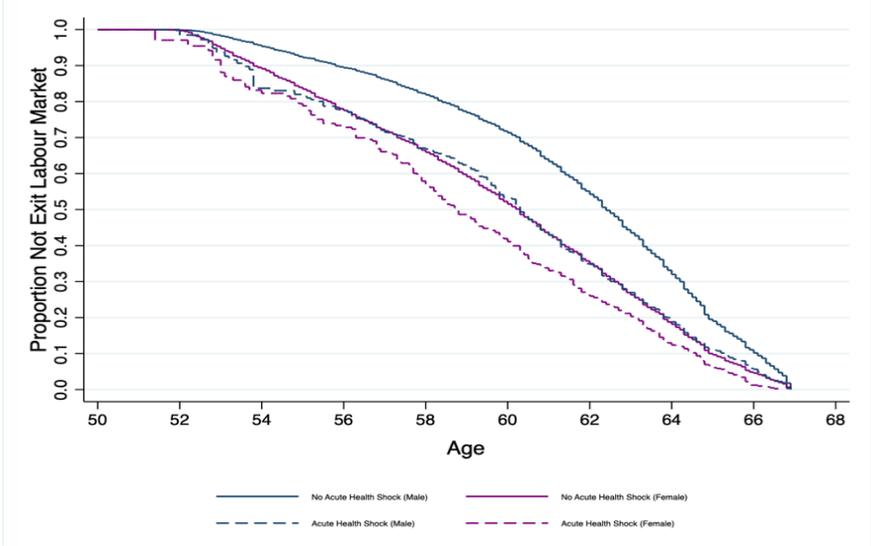

In Figure 5, the Kaplan-Meier survival curves illustrate that active individuals with fair or poor self-perceived health (SPH) have a significantly higher likelihood of exiting the labour market earlier compared to those reporting good, very good, or excellent health. By age 61, only around 20% of individuals with poor SPH remain active in the labour market, whereas approximately 40% of those with fair SPH still participate. In contrast, roughly 70-75% of individuals with very good or excellent SPH continue to be active. This large difference emphasises the crucial role of perceived health in influencing the timing of labour market exits.

Figure 5 – Kaplan-Meier Survival Estimates of the Proportion of Labour Market Exits Stratified by SPH (all countries)

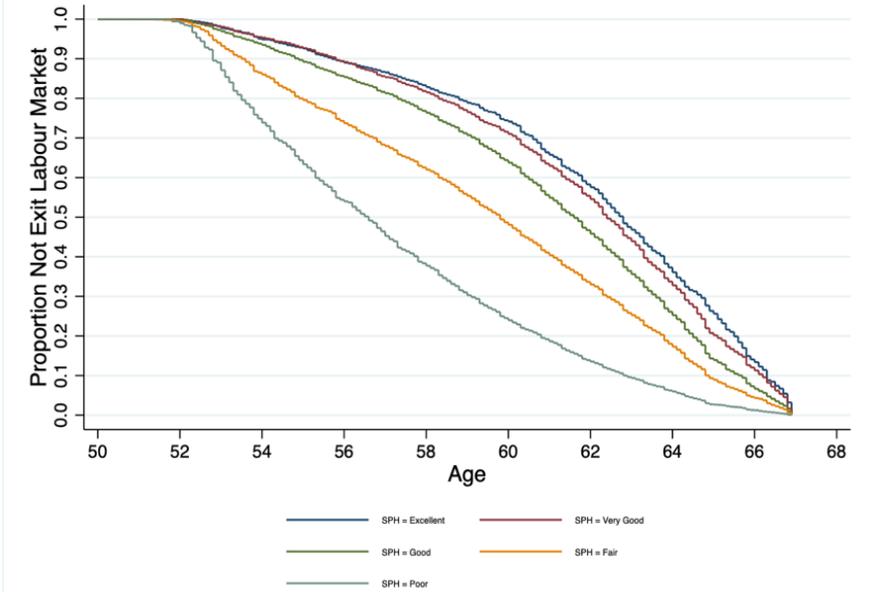

The necessity for a more nuanced analysis becomes evident when considering the previous figures referred to all countries, not accounting for potential country-specific differences in labour market exits and the role that acute health shocks, SPH and gender may play across various national contexts. I selected Belgium, Italy and Spain given these had larger number of observations and likely better represent the reality. They also display curious differences we will explore hereafter.

Figure 6 – Kaplan-Meier Survival Estimates Stratified by Acute Health Shock and Gender for Belgium, Italy and Spain

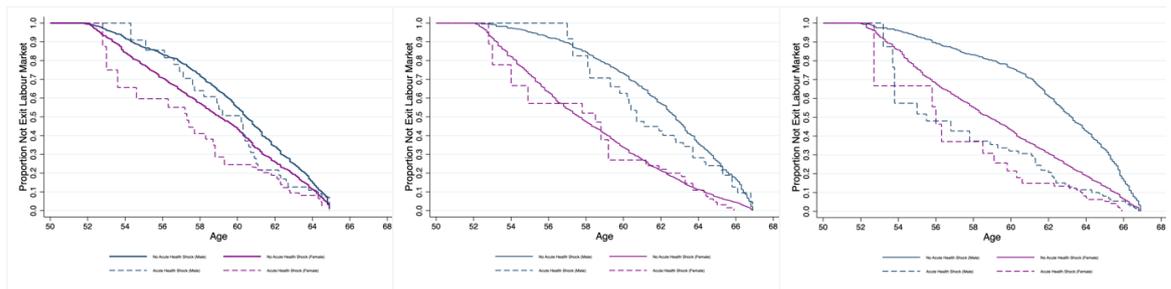

In Figure 6, the Kaplan-Meier survival estimates for Belgium, Italy, and Spain highlight distinct country-specific dynamics in labour market exits. Belgium shows a relatively small gender gap, with the impact of acute health shocks on labour exits being less pronounced for Belgian males and both Italian genders. However, acute health shocks significantly affect labour market exits among Belgian females, and the impact is especially pronounced among Spanish males. Italian females tend to exit the labour market more rapidly, while both Italian and Spanish males exhibit lower probabilities of early exits.

Figure 7 – Kaplan-Meier Survival Estimates Stratified by SPH and Gender Male for Belgium, Italy and Spain

| Belgium | Italy | Spain |

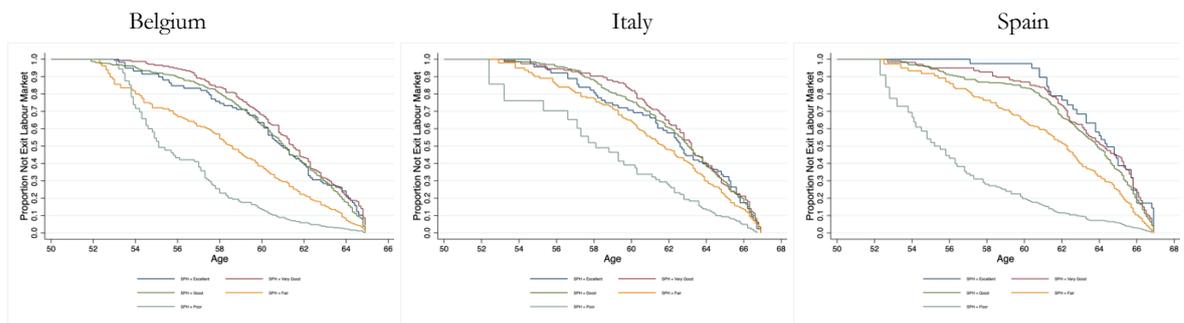

Across Belgium, Italy, and Spain, males with good to excellent SPH display (Figure 7) similar probabilities of remaining in the labour market over time, suggesting a consistent resilience among healthier individuals. Notably, in Italy, even those with Fair SPH display exit probabilities not markedly different from higher health strata, and the gap between Poor SPH and other groups is relatively narrow. In contrast, Belgium and Spain reveal dramatic different patterns for males with poor SPH, where both groups experience a steep and rapid decline in the likelihood of staying in the labour market.

Figure 8 – Kaplan-Meier Survival Estimates Stratified by SPH and Gender Female for Belgium, Italy and Spain

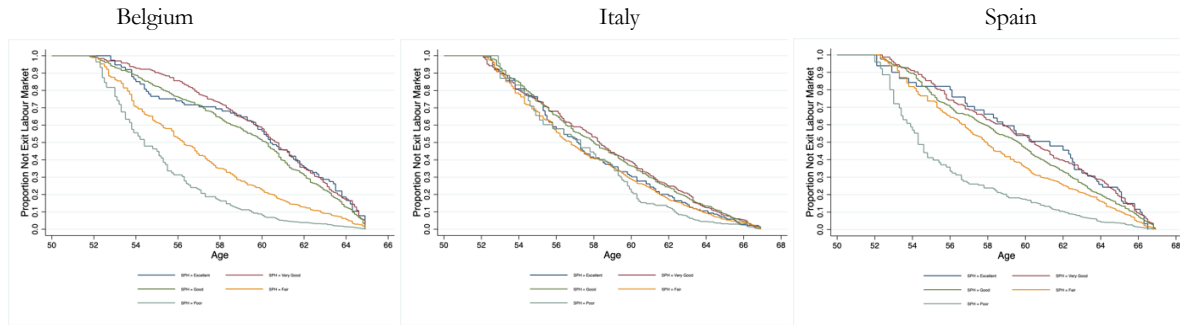

With regards to females, the Kaplan-Meier curves illustrate varying impacts of SPH on labour market exit probabilities. In Belgium, the differences between SPH levels are similar to the males, relatively less pronounced among higher SPH strata. While with fair and poor SPH leads to a more rapid decline. Italy shows a more consistent trend across all SPH levels, with only minor deviations for poor SPH after 60. Spanish females, like Belgium, exhibit a small difference between good to excellent SPH and those with poor and to some extent fair SPH too, reflect an accelerated exit from the labour market.

These variations underline the complexity of factors influencing labour market exits, where both acute health shocks and self-perceived health demonstrate varying impacts across Belgium, Italy, and Spain. These patterns may be influenced by gender-specific economic pressures, healthcare accessibility, and cultural norms around work and caregiving roles, which vary significantly across these countries. These country-specific patterns necessitate a more nuanced, multivariate approach to understand the interplay of health, economic, and cultural factors. Therefore, the Cox proportional hazards regression provides an essential framework, incorporating a broader set of covariates to comprehensively assess the risk of early labour market exit.

## 3.2 Cox Proportional Hazards

The Cox proportional hazards regression model provides a deeper view of the drivers behind early labour market exits, generally aligning with the patterns observed in the Kaplan-Meier survival estimates but offering a better understanding of the underlying factors driving early labour market exits. The hazard ratio represents the relative likelihood of an event occurring (in this case, labour market exit) at a point in time for individuals with a particular characteristic compared to those without it.

Table 3: Cox Hazard Regression Model for Labour Market Exits for Males

| Cox regression with Breslow method for the ties | |
|---|---|
| No. of subjects = 17,587 | Number of obs = 25,736 |
| No. of failures = 8,223 | |
| Time at risk = 69,886.2 | Wald chi2(14) = 1,002.96 |
| Log pseudolikelihood = -179,878.07 | Prob > chi2 = 0.0000 |
| | (Std. err. adjusted for 17,587 clusters in panelID) |

| _t | Hazard ratio | Robust std. error | z | P>z | [95% conf. | interval] |
|---|---|---|---|---|---|---|
| SPH (Reference: Excellent) | | | | | | |
|   Very Good | 1.008 | 0.046 | 0.19 | 0.853 | 0.923 | 1.102 |
|   Good | 1.233 | 0.053 | 4.83 | 0.000 | 1.132 | 1.342 |
|   Fair | 1.536 | 0.071 | 9.25 | 0.000 | 1.402 | 1.682 |
|   Poor | 2.444 | 0.133 | 16.45 | 0.000 | 2.197 | 2.718 |
| neg_HShocks | 1.152 | 0.029 | 5.61 | 0.000 | 1.097 | 1.211 |
| log_inc_m | 1.027 | 0.010 | 2.64 | 0.008 | 1.007 | 1.048 |
| ends_notmet | 1.057 | 0.025 | 2.35 | 0.019 | 1.009 | 1.106 |
| education_level (Ref: Univ. Degree) | | | | | | |
|   Diploma/Certificate | 1.132 | 0.064 | 2.18 | 0.029 | 1.012 | 1.265 |
|   Secondary Education | 1.373 | 0.039 | 11.13 | 0.000 | 1.298 | 1.451 |
|   Basic Education | 1.344 | 0.041 | 9.70 | 0.000 | 1.266 | 1.426 |
|   Other Education | 1.271 | 0.103 | 2.96 | 0.003 | 1.085 | 1.490 |
| 1.partner_hh | 0.769 | 0.043 | -4.69 | 0.000 | 0.690 | 0.859 |
| hh_comp (Ref: Solo-person) | | | | | | |
|   Two-person Household | 1.178 | 0.073 | 2.63 | 0.008 | 1.043 | 1.330 |
|   Three or more persons Household | 1.167 | 0.075 | 2.40 | 0.017 | 1.028 | 1.324 |

The Cox proportional hazards model for males in Table 3, reveals that most ratios are highly significant. SPH hazard ratios indicate that "Good" health increases exit risk by 23.3%, "Fair" health by 53.6%, and "Poor" health by 144.4% compared to "Excellent" health (reference category). Experiencing cumulative health shocks (neg_HShocks) raises the exit risk by 15.2%, while financial strain (ends_notmet) increases it by 5.7%, and slight 2.7% probability of exit increase for a log monthly income (log_inc_m) increase.

Education level impacts exit risk significantly, having a Diploma/Certificate increases the hazard by 13.2%, Secondary Education by 37.3%, Basic Education by 34.4%, and Other Education by 27.1%, relative to those with a University Degree. Living with a partner reduces the risk by 23.1%, whereas household composition influences exit risk, with individuals in two-person households showing a 17.8% higher risk and those in three-or-more-person households a 16.7% higher risk than those in solo households.

Overall for males, poorer health, lower education, and experiencing health shocks increases the likelihood of labour market exits, while living with a partner appears to reduce. Household composition also affects exit risk, with individuals in larger households showing higher hazards.

Table 4: Cox Hazard Regression Model for Labour Market Exits for Females

| Cox regression with Breslow method for the ties | |
|---|---|
| No. of subjects = 21,649 | Number of obs = 30,318 |
| No. of failures = 12,006 | |
| Time at risk = 81,349.8 | Wald chi2(14) = 1,605.88 |
| Log pseudolikelihood = -99,477.15 | Prob > chi2 = 0.0000 |
| | (Std. err. adjusted for 17,587 clusters in panelID) |

| _t | Hazard ratio | Robust std. error | z | P>z | [95% conf. | interval] |
|---|---|---|---|---|---|---|
| SPH (Reference: Excellent) | | | | | | |
| Very Good | 1.082 | 0.038 | 2.23 | 0.025 | 1.010 | 1.160 |
| Good | 1.205 | 0.041 | 5.48 | 0.000 | 1.127 | 1.288 |
| Fair | 1.344 | 0.049 | 8.07 | 0.000 | 1.251 | 1.444 |
| Poor | 1.868 | 0.084 | 13.92 | 0.000 | 1.710 | 2.040 |
| neg_HShocks | 1.054 | 0.021 | 2.66 | 0.008 | 1.014 | 1.096 |
| log_inc_m | 0.941 | 0.009 | -6.43 | 0.000 | 0.924 | 0.959 |
| ends_notmet | 1.137 | 0.021 | 6.84 | 0.000 | 1.096 | 1.179 |
| education_level (Ref: Univ. Degree) | | | | | | |
| Diploma/Certificate | 0.970 | 0.046 | -0.65 | 0.515 | 0.884 | 1.064 |
| Secondary Education | 1.453 | 0.035 | 15.57 | 0.000 | 1.386 | 1.523 |
| Basic Education | 1.535 | 0.038 | 17.41 | 0.000 | 1.462 | 1.610 |
| Other Education | 1.550 | 0.114 | 5.97 | 0.000 | 1.342 | 1.789 |
| 1.partner_hh | 1.212 | 0.044 | 5.30 | 0.000 | 1.129 | 1.301 |
| hh_comp (Ref: Solo-person) | | | | | | |
| Two-person Household | 1.011 | 0.041 | 0.28 | 0.778 | 0.935 | 1.094 |
| Three or more persons Household | 1.283 | 0.054 | 5.93 | 0.000 | 1.181 | 1.392 |

The Cox proportional hazards regression model for females (Table 4) shows several significant factors affecting labour market exits. Self-Perceived Health has a clear gradient effect, with "Very Good" health increasing exit risk by 8.2%, "Good" by 20.5%, "Fair" by 34.4%, and "Poor" health by 86.8% compared to the "Excellent" reference category. Cumulative health shocks (neg_HShocks) raise the exit risk by 5.4%, while financial strain (ends_notmet) increases it by 13.7%. Interestingly, income (log_inc_m) has a protective effect, with a 5.9% reduction in exit risk for unit increases in income.

For education, "Secondary Education" raises exit risk by 45.3%, "Basic Education" by 53.5%, and "Other Education" by 55% relative to those with a University Degree. Living with a partner (partner_hh) increases exit risk by 21.2%, while household composition has mixed effects: being in a two-person household does not significantly alter exit risk, whereas being in a three-or-more-person household increases it by 28.3%.

Comparing males and females, several patterns emerge. While poorer health consistently increases exit risk for both genders, females exhibit a lower hazard ratio across health levels, indicating that health impacts males more strongly. Income provides a protective effect for females, contrasting with its slight risk increase for males. Additionally, living with a partner reduces risk for males but increases it for females, suggesting divergent social or economic dynamics between genders in household dependency. Overall, education levels show a similar positive association with exit risk for both, but females in larger households face higher exit risks than their male counterparts.

# 4 Discussion & Conclusions

The analysis highlights significant factors influencing early labour market exits among individuals aged 50 and above, with acute health shocks considered as a crucial driver. Findings indicate that poorer self-perceived health (SPH), lower educational attainment, cumulative health shocks, and financial strain all play substantial roles in early retirement, though with gender-specific patterns.

For both men and women, deteriorating health status substantially increases the likelihood of labour market exit. Poor SPH, in particular, is associated with a clear high risk of early exit, suggesting that health is a primary determinant of continued employment. Financial factors also impact this risk; while financial strain modestly raises exit probability for both genders, higher income serves as a protective factor for women. Lower educational level is consistently associated with higher exit rates, aligning with the idea that individuals with fewer qualifications may have limited employment opportunities, especially if health deteriorates.
Gender differences emerge in the effects of household factors. While living with a partner reduces the risk of exit for men, it has the opposite effect for women, potentially reflecting different social roles or support dynamics. Similarly, larger household composition is associated with an increased exit risk for women, possibly due to greater familial responsibilities.

In light of these findings, targeted interventions addressing health and socioeconomic factors could delay early retirement. Policies aimed at supporting those in poor health or with limited education may help individuals remain in the workforce longer. Health programmes, retraining initiatives, and financial support services tailored to older workers, particularly women, may mitigate the pressures leading to premature labour market exits. Additionally, gender-specific support measures could better address the unique challenges that men and women face, especially in balancing work and household responsibilities. Promoting health maintenance and financial stability for older workers, through both preventive and supportive measures, may ultimately benefit individuals and broader social systems by reducing premature exits from the labour force.